\documentclass[12pt]{article}
\usepackage{amsmath}
\usepackage{amssymb}
\usepackage{mathtext}
\usepackage{graphicx}

\begin{document}
\section*{DYNAMICS OF THREE VORTICES\\ ON A PLANE AND A SPHERE --- II.\\
General compact case}\footnote{REGULAR AND CHAOTIC DYNAMICS, V.3, No.~2, \ 1998\\
{\it Received August 15, 1998}\\
AMS MSC 76C05}

\begin{centering}
A.\,V.\,BORISOV\\
Faculty of Mechanics and Mathematics\\
Department of Theoretical Mechanics
Moscow State University\\
Vorob'ievy gory, 119899 Moscow, Russia\\
E-mail: borisov@uni.udm.ru\\
V.\,G.\,LEBEDEV\\
Physical Faculty,\\
Department of Theoretical Physics
Udmurt State University \\
Universitetskaya, 1, Izhevsk, Russia, 426034 \\
E-mail: lvg@uni.udm.ru\\
\end{centering}

\begin{abstract}
Integrable problem of three vortices on a plane and sphere are considered.
The classification of Poisson structures is carried out.
We accomplish the bifurcational analysis using the variables introduced in
previous part of the work.
\end{abstract}

\section{Introduction}

In this part of our work the integrable problem of dynamic of vortices
on  planes and sphere are considered. We base on Hamiltonian form of reduced
systems, which poisson structure was given in the previous part \cite{BP}.

The problem of motion of two vortices on plane was completely
investigated by G.\,Helmholz \cite{Melesh}, who had established,
that in general case two vortices
make an uniform rotary motion around of vorticity centre, with frequency
$$\Omega = \frac{\Gamma _ 1 + \Gamma _ 2}{2\pi r ^ 2}, $$
where $ \Gamma _ k $ is intensity of point vortices.
The vorticity  centre thus is moving uniformly and rectilinearly.
If  $ \Gamma _ 1 = -\Gamma _ 2 $
the vorticity centre is situated on infinity, and two vortices are moving
forward.

The problem about a motion of three vortices is much more complex. Despite of
numerous works, first of which were the dissertation of ~Gr{\"o}bli 1877
\cite{Grobli} and Greenhill's research \cite{green}, the complete and evident
classification of a motion still does not exist.
From the point of view of integrability problem, this
problem was considered by Poincare in the treatise on the vortices theory
\cite{Poincare}. He wrote out explicitly a complete set of noncommutative
integrals.
In a modern period the problem of a motion of three vortices on a plane was
studied in works \cite{Aref,Aref2,Rott,TT,Novikov},
and from the point of view of the topological analysis --- in
\cite{Selivanova}. Unfortunately, these works have added a little to
achievements of classicists both
in presentation, and in completeness of the description of motions. Their
basic contents come down either to geometrical interpretation of a
motion and computer modeling of separate trajectories, or to some general
topological constractions, that is not coordinated to physical behaviour of
system. Partly, it is caused by the fact, that the problem of three vortices on a
plane does not belong to that integrable systems, which complete analysis is
possible in a class of rather simple (for example, elliptic) special functions
(the general solution  has indefinitely--sheeted branching on complex plane
of time because of logarithmic terms in Hamiltonian). The exception consists
of some special cases (for example, the case of equal intensities of
vortices). In comparison with a problem of three vortices on a plane, a
motion of three vortices on sphere, that is also being integrable, is not
investigated at all.

We give here new analysis, based on algebraic and geometrical research of
the reduced system (in variable, which in the previous part of work \cite{BP}
were named ``internal'',  as though we are not adhere this term anymore),
and then we analyze absolute motion, without use of exact quadrature. Such
approach, basing on representation of the equations of motions on algebra
\cite{BP}, allows to establish some analogies between problem of three
vortices and Euler-Poinsot case in rigid body dynamics, and also to receive
more evident description of motions of system.

\section{Algebraic classification}

As it was shown in the previous part of work \cite{BP}, the equation of a
motion of three vortices on the plane can be written down as Hamiltonian system,
determined by Lie-Poisson brackets in form of
\begin{equation}
\label{Tpb}
\begin{array}{l}
\displaystyle \{M _ i, M _ j \} =-4a _ k\Delta,\medskip\\
\displaystyle \{M _ i, \Delta \} = (a _ j -a _ k) M _ i +
                             (a _ j + a _ k) (M _ j-M _ k), \\
\end{array}
\end{equation}
and  by Hamiltonian
\vspace{-4mm}
\begin{equation}
\label{Tpa}
H = -{1\over 8\pi}(\Gamma _{2} \Gamma _{3} \log M _{1} + \Gamma _{1}
\Gamma _{3} \log M_{2} +
\Gamma _{1} \Gamma _{2} \log M _{3}).
\end{equation}
where $M_k$ are squares of pairwise distances between vortices,
$\Delta$ is a  size of the oriented area pulled on three vortices
(here and further we shall assume, that the indexes $ i, j, k $ accept
accordingly values $ 1,2,3 $ and their cyclic rearrangements,   and the
coefficients $ a _ 1, a _ 2, a _ 3 $ are inverse intensities $ a _ k = 1 /
{\Gamma _ k}). $

Lie-Poisson brackets (\ref{Tpb}) are degenerated and have two central functions.
One of them (linear) is integral of the complete moment
\vspace{-3mm}
\begin{equation}
\label{Tpá}
D = \sum _ k a _ k M _ k,
\end{equation}
the other (square-law Casimir function) arises from geometrical
Henon ratio, that connects the area of triangle with sides
\begin{equation}
\label{Tpd}
F = (2\Delta) ^ 2 + M _ 1M _ 1 + M _ 2M _ 2 + M _ 3M _ 3-2 (M _ 1M _ 2 + M _ 1M
_ 3 + M _ 2M _ 3).
\end{equation}
For real motions $ F = 0. $

The real type of Lie-Poisson algebra (\ref{Tpb}) depends on values of
intensities $ \Gamma _ 1, \Gamma _ 2, \Gamma _ 3. $
In fact, choosing a new basis in form $D,e_1,e_2,e_3$, where $D$ is
defined by~(\ref{Tpá})
\begin{equation}
\label{Tp2}
\begin{array}{l}
\displaystyle e _ 1 = \frac{\Delta}{2\sqrt {A}}, \\
\displaystyle e _ 2 = \frac {(a _ 2-a _ 3) M _ 1 + (a _ 3-a _ 1) M _ 2 + (a _ 1-
a _ 2) M _ 3}{2\sqrt {2AB}}, \\
\displaystyle e _ 3 = \frac
              {(a _ 2a _ 3-a _ 1 ^ 2) M _ 1 + (a _ 1a _ 3-a _ 2 ^ 2) M _ 2 + (a
_ 1a _ 2-a _ 3 ^ 2) M _ 3}{2A\sqrt {2B}} \\
\end{array}
\end{equation}
where
$$A=|a_1a_2+a_2a_3+a_1a_3|,$$
$$B=(a_1-a_2)^2+(a_2-a_3)^2+(a_1-a_3)^2,$$
under the condition
\vspace{-4mm}
\begin{equation}
\label{Tp5}
(a_1a_2+a_2a_3+a_1a_3)>0,
\end{equation}
we receive that the algebra of vortices is decomposed in the direct sum
$l(4)\approx$ $\mathbb{R}\oplus so(3):$
\begin{equation}
\label{Tp7}
\{D,e_k\}=0\mbox{, }
\{e_1,e_2\}=e_3\mbox{, }
\{e_2,e_3\}=e_1\mbox{ }
\{e_3,e_1\}=e_2,
\end{equation}
and by condition
\vspace{-4mm}
\begin{equation}
\label{Tp6}
(a_1a_2+a_2a_3+a_1a_3)<0,
\end{equation}
in direct sum $l(4)\approx$ $\mathbb{R}\oplus so(2,1):$
\begin{equation}
\label{Tp8}
\{D,e_k\}=0,\quad \{e_1,e_2\}=-e_3,\quad
\{e_2,e_3\}=e_1,\quad
\{e_3,e_1\}=e_2.
\end{equation}
Though for case equal intensities the coefficient $B=0$,
using the passage to the limit, it is easy to show that basis
(\ref{Tp2}) can be correctly determined in this case as well.

Let's consider, what features of a motion can be obtained from the structure
of such algebraic decomposition. Using squared Casimir function for
algebra $ so(3) $
\begin{equation}
\label{Tp9}
G ^ 2 = e _ 1 ^ 2 + e _ 2 ^ 2 + e _ 3 ^ 2,
\end{equation}
and expressing in it a vector $ e _ 1 $ with the help of Heron ratio
(\ref{Tpd})
with $ F = 0, $ we discover, that relative dynamics of vortices with
intensities, satisfying to a condition (\ref{Tp5}) will be equivalent to a
motion of some "representing" point on symplectic sheet, that is a
two-dimensional sphere with radius, determined by value of linear Casimir
function
\begin{equation}
\label{Tp10}
e _ 1 ^ 2 + e _ 2 ^ 2 + e _ 3 ^ 2 =\frac 1 {16}
\left (\frac {a_1a_2a_3 D}{a _ 1a _
2 + a _ 2a _ 3 + a _ 1a _ 3} \right
) ^ 2.
\end{equation}
The situation  for algebra $ so (2,1) $ is similar that arise
in a case (\ref{Tp6}). Using Casimir function of algebra~$so(2,1)$
\begin{equation}
\label{Tp17}
G ^ 2 = e _ 3 ^ 2-e _ 2 ^ 2-e _ 2 ^ 2,
\end{equation}
it is possible to notice, that the relative motion of vortices is reduced
to the motion of point on surface of two-cavities hyperboloid, that is
determined by condition,
\begin{equation}
\label{Tp18}
e _ 3 ^ 2-e _ 1 ^ 2-e _ 2 ^ 2 =\frac 1 {16}
\left (\frac {a_1a_2a_3D}{a _ 1a _ 2 +
a _ 2a _ 3 + a _ 1a _ 3} \right
) ^ 2.
\end{equation}
In case of (\ref{Tp5}), motions of representing point (and therefore
three vortices in system of vorticity centre) is finite for every $D$,
therefore this case we shall futher name ``compact'',
and in case (\ref {Tp6}), with which obviously may be exist running up
trajectories by additional conditions for $\Gamma_i$ and for initial
vortices positions, the name ``uncompact''.

Motions, arising under condition
\begin{equation}
\label{Tp20}
( a _ 1a _ 2 + a _ 2a _ 3 + a _ 1a _ 3) = 0
\end{equation}
require separate consideration. In basis
\begin{equation}
\label{Tp23}
\begin{array}{l}
\displaystyle D=\frac {a_1M_1+a_2M_2+a_3M_3}{(a_1+a_2)^2},\\
\displaystyle e_1=\frac {\Delta}{(a_1+a_2)},\\
\displaystyle e_2=\frac {M_3}{2(a_1+a_2)},\\
\displaystyle e_3=\frac {(a_1+a_2+a_3)M_3-(M_1+M_2)(a_1+a_2)}{2(a_1+a_2)^2},\\
\end{array}
\end{equation}

Lie-Poisson brackets can be reduced  to form
\begin{equation}
\label{Tp27}
\displaystyle \{ D,e_k\}=0,\quad \displaystyle \{ e_1,e_2\}=e_3,\quad
\displaystyle \{ e_3,e_1\}=-D,\quad \displaystyle \{ e_2,e_3\}=e_1.
\end{equation}
In this case algebra of vortices is not decomposed in the straight sum of
one-dimensional and three-dimensional algebras, and becomes four-dimensional
solveable algebra with maximal solveable ideal $N = \{D, e _ 1, e _ 3 \}.$
Algebra has squared Casimir function, following from (\ref{Tpd})
\begin{equation}
\label{Tp28}
G ^ 2 = e _ 1 ^ 2 + e _ 3 ^ 2-2De _ 2.
\end{equation}
Real dynamics of vortices occurs on symplectic sheet,
that is set by conditions $G = 0$ and $D~=~const,$ which
determine paraboloid, taking place through the beginning of coordinates.
By $ D = 0 $, being necessary collaps condition (merge of vortices),
paraboloid degenerates to the straight line, con\-ter\-mi\-nous with an axis
$ e _ 3.$

\section{Bifurcational analysis of motion of vortices on plane}

Let's describe most evident geometrical interpretation of motions, used in
\cite{TT} and presented on~Fig.~1. In space
$ M _ 1, M _ 2, M _ 3 $ the level of linear integral of the moment sets a
plane.
The inequalities $ M _ k> 0 $ allocate on it  the area, in which motion
occurs. It is
necessary to exclude from this area unphysical values of  distances, for
which an
inequality of a triangle $ (\Delta^ 2 < 0)$ does not hold. This area is shown
at
a figure by black colour. When approaching to it, in the equations of a
motion
for variable $M _ 1, M _ 2, M _ 3, $ we schould change the sign of time and
direction of motion. It is necessary to note, that
the described geometrical interpretation is not equivalent to a motion of
representing point on symplectic sheet, and changes of direction of motions
are the consequence of features of a projection from space
$ \Delta, M _ 1, M _ 2, M
_ 3 $ in space $ M _ 1, M _ 2, M _ 3. $

Let's hold condition (\ref{Tp5}), then bracket (\ref{Tpb}) is defines algebra
$so(3)\oplus \mathbb{R}$. In this case symplectic sheet is compact ($\mathbb{S}^2$), and
motions of vortices are finite. Let's assume, at first, that signs of all
intensities are identical --- in this case condition (\ref{Tp5}) is obviously
fullfiled. The first integrals of a motion of vortices on a plane we shall
write down as

\begin{equation}
\label{Tp11}
\begin{array}{l}
\displaystyle D = a_1 M_1 + a_2 M_2 + a_3 M_3,\medskip\\
\displaystyle h = M_1^{a_1} M_2^{a_2} M_3^{a_3}, \\
\end{array}
\end{equation}
(integral of energy, for convenience, is presented in
exponential form). The integrals (\ref{Tp11}) are dependent, i.e.
Jacobi matrix of the first integrals (\ref{Tp11}) is degenerate
\begin{equation}
\label{Tp12}
Rank \left (\frac {\partial (D, h)}{\partial (M _ 1, M _ 2, M _ 3)} \right) < 2
\end{equation}
only in one case: $ M _ 1 = M _ 2 = M _ 3 $
(in three-dimensional space $ M _ 1, M _ 2, M _ 3 $ the vortices will form
correct triangle). Curves, resulting with crossing
of levels of integrals (\ref{Tp11}) are similar to polhodes
in rigid body dynamics (Poinsot interpretation of  Euler case)
and, in a compact case, are ovals  (Fig.~1). The solution as correct
triangle occurs with a contact of a surface $ h $ and plane $ D $
in one point. They limit by energy the area  of  possible motions (APM) from
above and the corresponding bifurcational curve has a form
\vspace{-5mm}
\begin{equation}
\label{Types13}
H (D) = \left (\frac {D}{\sum a _ i}\right)^{(\sum a_i)}.
\end{equation}
{\bf Remark 1.} {\it The particular solutions, appropriate
to the given curve --- three vortices in tops of correct triangle,
rotating as a rigid body around of vorticity centre are refered as
``Thomson's" and are steady with fulfilment of a condition (\ref{Tp5}). J.
J. Thomson has specified them for any number of vortices of equal
intensity, and has shown, that in linear approach such configurations will
be steady for number of vortices $ N\leqslant 6 $, and with $ N\geqslant 7
$ --- unstable (Thomson theorem) \cite{ThomsonJ}. However because of
presence of resonances in system the linear approach is not sufficient.
The analysis of stability in nonlinear approach  with use of  Birkhoff
normalization was carried out in \cite{Hazin}. It was found, that the
theorem Thomson is fair in the exact statement --- in sense of  Laypunov
stability.}

The other change of motion occurs with a contact of a crossing
line of integrals' levels with curves, determined by the equation
$\Delta (M_1, M_2, M_3) = 0$ $(\sqrt {M_i} + \sqrt {M_j} = \sqrt {M_k}),$
that arise from inequalities of triangles and limit physically
allowable area of values (see Fig.~1).

Points of a contact correspond to collinear configuration of three vortices.
Three vortices thus settle down on one straight line and rotate as a unit
around of vorticity centre.

{\bf Remark 2.} {\it Collinear and the triangular configurations in dynamics of three
vortices have
analogues in the classical selestial mechanics \cite{ArnoldKozlovNei}. There
are, accordingly, Euler and Lagrange particular solutions of three body
problem.}

Collinear configurations are obtained from a condition of a contact
\begin{equation}
\label{Tp13}
\Delta = 0, \quad \dot \Delta = 0.
\end{equation}
that lead, having same sedate functional dependence,
which in a general case can be presented as
\begin{equation}
\label{Tp14}
H(D)=f(a_1, a_2, a_3)D^{(a_1+a_2+a_3)},
\end{equation}
where $f(a_1,a_2,a_3)$ some function from parameters.
Passing to homogeneous coordinates $ x, y $
\begin{equation}
\label{Tp15}
M_1 = xM_3,\quad  M_ 2 = yM _ 3,\quad y = (1 + x) ^ 2, \quad x = z ^ 2,
\end{equation}
we discover, that collinear configurations correspond to positive
roots of the equations of the third degree
\begin{equation}
\label{Tp16}
\alpha_1 (1\pm z) (z\pm 2) + \alpha _2 ((1\mp z) z-\alpha _3 (1\pm 2z) z (1\pm
z) = 0,
\end{equation}
where
$$\alpha _1 ={\Gamma _2 + \Gamma _3}, \quad
\alpha _2 ={\Gamma _1 + \Gamma _3}, \quad
\alpha _3 ={\Gamma _1 + \Gamma _2}. $$
The lines of crossing of levels $ D $ and $ h $ can concern borders of area
of physically possible motion $ \Delta = 0 $ in space of variables
$ (M _ 1, M _ 2, M _ 3) $ by various ways (Fig.~1). Depending on it the
quantity of bifurcational curves also will be different. In particular, in a
case, when all of intensities are different $ a _1 \ne a _2\ne a _3\ne a _1 $
three branches corresponding to collinear motions will exist (Fig.~2a).
With equality of two of them, for example, $ a _ 1 = a _ 2\ne a _ 3 $
two appropriate bifurcational curves merge. And, at last, in case of equal
intensities $ a _ 1 = a _ 2 = a _ 3 $ all three curves degenerate in one.

The calculation of angular speed of Thomson's and  collinear configurations
concerning vorticity centre is given, for example in \cite{Melesh}.  It
monotonously decreases with increase of the complete moment of system
vortices.

As a measure of stability of stationary configurations it is possible
to use a square of untrivial eigenvalues $\lambda^2$ of
linearized system of the equations of a motion of three vortices. For
Thomson's configurations it is simple to receive the explicit formula
\begin{equation}
\label{Tpf}
\lambda ^ 2 = -3a _ 1a _ 2a _ 3\frac {(a _ 1 + a _ 2 + a _ 3) ^ 3}{D ^ 2},
\end{equation}
that shows, that in a considered case they are steady, in
difference from collinear, which as shown on Fig. 2a a are unstable.

Generally speaking, the following statement is fair, that is confirmed
by geometrical interpretation on Fig.~1:
If Thomson's solution is steady, the appropriate set
of collinear solutions  are unstable, and on the contrary.
Therefore condition (\ref {Tpf}) determines a type of
stability not only Thomson's configurations, but also collinear
configurations with the given values of $D.$

Let's consider other case of a motion of three vortices,
with fullfiled conditions (\ref{Tp5}) (easy to see, that all other possible
cases, when the condition is fair, could be reduced to two considered cases).
Let's assume, that the intensity of one of vortices has opposite sign in
comparison with other two (for example, $ \Gamma _ 1 = 1/a _ 1 < 0 $). The
condition (\ref{Tp5}) in this case means, that $ -\Gamma _ 1 < \Gamma _ 2 +
\Gamma _ 3 $ (i.e. intensity of  chosen vortex is greater than the intensity
of other two).

Let's specify only main differences of this problem from previous. First of
all from the formula~(\ref{Tpf}) for Thomson's configurations it follows,
that they are unstable. Collinear configuration (it is not three of them
anymore, but only one) thus, on the contrary, is steady. Bifurcational curve
of appropriate unique collinear configuration (see Fig.~2b) is above a
curve, appropriate to Thomson's configuration (it is simple to show, that
under condition of~(\ref{Tp5}) complete moment is always more than
zero), and the area of possible motion coincides with the whole square
$ D> 0, \quad h> 0 $.

To obtain more complete representation about the considered motions, let's
study a phase portrait of system of three vortices on a plane in symplectic
coordinates.

\section{Symplectic coordinates for vortices on a plane}

Since symplectic sheets represent two-dimensional spheres
(\ref{Tp11}), it is convenient to use cylindrical coordinates $ l, L $ as
canonical  coordinates (we shall keep for them a designation of Andoyer--Deprit
coordinates, adopted in rigid body dynamics  \cite{BorEm}) for algebra
$ so (3): $
\begin{equation}
\label{Tp18a}
\begin{array}{l}
\displaystyle e_1 = L, \\
\displaystyle e_2 = \sqrt {G^2-L^2} \sin l, \\
\displaystyle e_3 = \sqrt {G^2-L^2} \cos l. \\
\end{array}
\end{equation}
Expressions for variables $ M _ 1, M _ 2, M _ 3 $ in
canonical coordinates look
like
\begin{equation}
\label{Tp19}
\begin{array}{ll}
M_i=&\frac {\displaystyle a_j+a_k}{\displaystyle \sqrt {2}}{\displaystyle \frac
{D}{A}}+2F(a_j-a_k)F\sqrt {\frac {\displaystyle A}{\displaystyle B}}\sin l+\medskip\\
&{\displaystyle \frac{4F}{\sqrt{B}}}(a_j(a_j-a_i)+a_k(a_k-a_i))\sqrt{\frac{2}{B}} \cos l,\\
\end{array}
\end{equation}
where
$$ A = a _ 1a _ 2 + a _ 2a _ 3 + a _ 3a _ 1, $$
$$ B = (a _ 1-a _ 2) ^ 2 + (a _ 2-a _ 3) ^ 2 + (a _ 1-a _ 3) ^ 2, $$
$$ F = \sqrt {G ^ 2-L ^ 2}. $$

In that special case of equal intensities the derived expressions
(\ref{Tp17}) lead to the expression for Hamiltonian mentioned in \cite{Aref,
Khanin}. In last work it is used the classical way of a reduction to a
problem with one degree of freedom for arbitrary intensities (such as
Jacobi's unit elimination).

The phase portrait in canonical coordinates describes a motion
of representing point $ (L, l) $ on a surface of two-dimensional  sphere. The
pulse $L$ has a value of  the oriental area parallelogram, that is
constructed
on three vortices $(L = 2\Delta)$. Development of a phase portrait
is presented on Fig.~4a--d for four different specific cases of  the
intensities of  vortices ratio.

The given figures illustrate the analogy between the motion of three vortices
and rigid body dynamics. Comparing Fig. 2a  (in case of equal
intensities) with phase
portrait of a Euler-Poinsot problem (see for example \cite{BorEm}),
it is possible to connect  a collinear configuration (situated on a straight
line $ L = 0 $) with unstable permanent motions of a rigid body around of an
middle axis of ellipsoid of inertia, the Thomson's solutions (with $ L/G = 1$)
--- with rotations around of a large (small) axis of ellipsoid of inertia.
The periodic solutions of a problem of two vortices (two of three vortices
always are in one point, and their intensities are added), situated on
straight line $ L = 0 $ it is possible to connect with steady permanent
rotations around of a small (large) axis of elipsoid of inertia. When system
pass through  the collinear state ( three vortices on one straight line ---
$ L = 0 $), oriented area changes the sign. Deformation of a phase picture
with unequal intensities is shown on Fig.~4b,c. Thus Thomson's  solutions
are displaced with straight $ L/G = 1, $ that on sphere should be imagined as
being pulled together in a point). Fig.~4d present the phase portrait, with
one negative intensity, but condition (\ref{Tp5}) is also fullfiled.

Other motions of system of three vortices, that is not a state of equilibrium on
phase portraits, correspond to quasiperiodic motions of vortices on a plane.
Depending on the topology of a phase curve these motions have
various features --- some of them repeatedly pass
through the collinear state, others --- never get into it. Such areas
are separated by separatrix curve, for that curve three vortices are
approach to collinear (unstable) configuration infinitely long.
The particular trajectories of vortices are given in various works
\cite{Aref2, Aref, Rott, Melesh,TT}, beginning from Gr\"obli~\cite{Grobli}
and Greenhill \cite{green}.

Appropriate simplectic coordinates for algebra $ so (2,1) $ could be
chosen as
\begin{equation}
\label{Tp20a}
\begin{array}{l}
\displaystyle e _ 1 = L, \\
\displaystyle e _ 2 = \sqrt {G ^ 2 + L ^ 2} \mbox {\, sh}\,l, \\
\displaystyle e _ 3 = \sqrt {G ^ 2 + L ^ 2} \mbox {\, ch}\,l. \\
\end{array}
\end{equation}
where $G,$ as well as in a case $so(3)$, parametrize various
simplectic sheets. The difference of the case (\ref{Tp6})
is displayed in noncompactness of symplectic sheet, therefore
the leaving of vortices to infinity is possible, that is excluded for
condition (\ref{Tp5}).

\section{Three vortices on sphere}

The problem of motion of vortices on sphere is more interesting, but less
investigated. This problem has  more important applied meaning in comparison
with a problem of motion of vortices on a plane, because it is directly
connected with physics of atmosphere. Many substantial problems (collapse
and evasion of vortices, existence of steady stationary (static)
configurations)
can be investigated on model of an ideal liquid. Their solutions can become a
basis for more complex models (connected, for example, with presence of small
viscosity), that describe real dynamic vortical processes (motion of
cyclones and etc.) in an atmosphere of the Earth.

Dynamics of two vortices on sphere is quite similar to a flat case.
Here the general situation is the rotation around of an axis, that is passing
through centre of spheres and analogue of
vorticity centre (point located on chord, connecting two
vor\-te\-ces with radius-vector
$$\vec r = \frac {\Gamma _ 1\vec r _ 1 + \Gamma _ 2\vec r _ 2}{\Gamma _ 1 +
\Gamma _ 2}, $$
where $\vec r _ 1, \vec r _ 2 $ --- radius-vector of the first and second
vortices). As well as in a flat case distance between vortices remains
constant.
Angular speed of such rotation
\begin{equation}
\label{Sp111}
\Omega = \frac {1}{2\pi M} \sqrt {(\Gamma _ 1 + \Gamma _ 2) ^ 2-\frac {\Gamma
_ 1\Gamma _ 2M}{R ^ 2}},
\end{equation}
where $ R $ --- radius of sphere, $ M $ --- square of distance between the
vortices (square of the length of a chord, that connect the location of two vortices
on sphere). Under condition of $ \Gamma _ 1 = -\Gamma _ 2 $, two vortices are
moving on sphere on two identical parallels, located on different parties from
equator, that was noticed Gromeka \cite{Grom}.

The problem of motion of three vortices on sphere was considered in work
\cite{Bog1}, in that work probably for the first time the integrability of the
problem was specified.
The equations of motion of three vortices on sphere in
variables of mutual distances (chords) and volumes can be
written down~\cite{BP} as the Hamiltonian equations with nonlinear algebra of Poisson brackets
\begin{equation}
\label{Sp1}
\begin{array}{l}
\displaystyle \{M _ i, M _ j \} =-\frac {4a _ k\Delta} {R}, \\
\displaystyle \{M _ i,
\Delta \} = R (a _ j-a _ k) M _ i + R (a _ j + a _ k) (M _ j-M _ k) + \frac {M _
i}{2R} (a _ kM _ k-a _ jM _ j), \\
\end{array}
\end{equation}
that also have two central functions, one of
them (linear) is the same as (\ref{Tpá}), and the
second (cubic) function has the form
\begin{equation}
\label{Spd}
\begin{array}{rl}
F=&(2\Delta) ^ 2 + R^2(M _ 1M _ 1 + M _ 2M _ 2 + M _ 3M _ 3)-2 R^2(M _ 1M _ 2 +\medskip\\
&M _ 1M _ 3 + M _ 2M _ 3)+M_1M_2M_3 = 0,
\end{array}
\end{equation}
and Hamiltonian (\ref{Tpa}).

Let's notice, that linear approximation of structure (\ref{Sp1})
coincides with the Poisson bracket of three vortices on a plane. Besides,
it is simple to show, that after replacement
$$
\frac {d\tau}{dt} = \pm\frac {\Delta}{2\pi M _ 1M _ 2M _ 3}
$$
the equations of a motion of vortices turn in Lotky--Volterra system of the form
\begin{equation}
\label{Spl}
\frac d{d\tau} M_i=\Gamma_iM_i(M_j-M_k),
\end{equation}
similarly to the flat case \cite{BP}. This trajectory isomorphism is also
piecewise \cite{BP}.

Geometrical interpretation for a plane, presented on
Fig.~1, can be also transferred to sphere. Curious and
earlier, probably, not marked fact is that phase trajectories in variable
$M_1,M_2,M_3$ for a case of sphere and plane,
with given intensities, coincide with phase trajectories
of the same Lotky--Volterra system (\ref{Sp1}). Thus main effects
in dynamics of vortices are determined by the part of phase
trajectories of Lotky--Volterra system gets that in area
$\Delta^2(M_1,M_2,M_3)>0.$ The motion of vortices occurs only
in this part, because with the approach of trajectories to border of area
(and achievement by vortices collinear state) in the equations (\ref{Spl})
it is necessary to change a sign of time.
The form of this area is a little various
for a plane and sphere (formula (\ref {Tpd}) and (\ref {Spd}))
and represented on Fig.~7.

{\bf Remark 3.} {\it The analogy with Lotky--Volterra system is
very useful for study of vortices collaps problem and is related to
Bolin regularization in the  \cite{ArnoldKozlovNei} (the
difference is, that in the Kepler problem is fixed a constant of energy
level). Problems of regularization and collapse, that arise only in noncompact
cases, will be discussed in the next part of work.}

{\bf Remark 4.} {\it The issue of reduction of
three vortices problem on sphere to canonical Hamiltonian system
with one degree of freedom is complex and, probably, it could not be solved
constructively. If one use the initial form of record \cite{BP},
then one obtain the six-dimensional Hamiltonian system, that have as the
integrals Hamiltonian function and noncommutative set of integrals of moment,
each of them is nonlinear function of phase variable (in difference from a
case of a plane considered in \cite{Khanin}).}

{\it The canonization of the reduced system also problematic,
in variable $ (\vec M, \Delta) $ the canonization is equivalent to
introduction of Darboux coordinates on two-dimensional symplectic sheet
determined by a common level of Casimir functions. It require the the
solution of nonlinear system of the partial differential equations.}

{\it Really, the nonlinear algebra of Poisson brackets (\ref{Sp1}) cannot
be investigated in same detail, as in the flat case. Linear
approximation of this structure is capable to make some
qualitative conclusions only for a situation
close to simultaneous collaps (i.e. when distance between
vortices is small in relation to radius of curvature).
From it, nevertheless, it is possible to make a conclusion that
necessary conditions of simultaneous collaps of vortices on a plane
are also fair for the case of sphere, since the influence of nonlinear terms
is small enough near the collaps.}

{\it The most simple form the bracket (\ref{Sp1}) has in case of three equal
intensities at a zero level of linear Casimir function $ (D = 0). $
After transition (for linearization of a bracket (\ref{Sp1})) to
canonical basis (\ref{Tp2}) it is possible to present bracket}
\begin{equation}
\label{Sp22}
\begin{array}{l}
\displaystyle \{e _ 1, e _ 2 \} = e _ 3 + \frac {1}{2} (e _ 3 ^ 2-e _ 2 ^ 2),
\\
\displaystyle \{e _ 2, e _ 3 \} = e _ 1, \\
\displaystyle \{e _ 3, e _ 1 \} = e _ 2-e _ 2e _ 3. \\
\end{array}
\end{equation}
{\it It is curious to note, that nonlinear terms, included in a bracket,
(\ref{Sp22}), are similar to the terms, originating in Kovalevsky integral
of the Euler--Poisson equations \cite{BP}.}

{\it One of main sections of  sheet for a case
$ \Gamma _ 1 = \Gamma _ 2 = \Gamma _ 3 $ are shown on Fig.~5 in
dependence from curvature of sphere $ k = 1/R $.
With $ k = 0, $ the symplectic sheet, that is determined by algebra
$ So (3) $, is also sphere. The change $ k $ results in a surface, whose
coherent component is homeomorphic to sphere. It would be interesting to
investigate the topology of a symplectic sheet with various values of
intensities and values of integral of the moment $ D $, that is the limited
function on sphere (on which vortices move).
Last remark allows to reject uncompact
components of symplectic sheets, since a motion on sphere is always
finite.}

{\bf Remark 5.} {\it The general principle of the classification of vortices motions on sphere
given below is the continuation on parameter of complete moment of stationary
configurations that is known near $ D = 0. $ In last case the influence of
curvature is not significant, that corresponds already investigated
flat problem (it is also equivalent to consideration of linear
approximation of structure (\ref{Sp1})).}

The conditions for Thomson's of configurations on sphere are similar to
conditions on plane
$$ M _ 1 = M _ 2 = M _ 3. $$
Conditions for existence of collinear configurations
$ (\Delta = \dot \Delta = 0) $ in case of sphere are reduced
to system of three algebraic equations.
Introducing constants
$$\alpha _ 1 = \Gamma _ 3 + \Gamma _ 2, \quad \alpha _ 2 = \Gamma _ 1 + \Gamma _
3, \quad
\alpha _ 3 = \Gamma _ 2 + \Gamma _ 1,\quad
\beta _ 1 = \Gamma _ 3-\Gamma _ 2, \quad \beta _ 2 = \Gamma _ 1-\Gamma _ 3,
\quad
\beta _ 3 = \Gamma _ 2-\Gamma _ 1, $$
we write down system of the equations as
\begin{equation}
\label{Vr14}
\begin{array}{l}
\displaystyle 2 (M _ 1M _ 2 + M _ 1M _ 3 + M _ 2M _ 3) - (M _ 1 ^ 2 + M _ 2 ^ 2
+ M _ 3 ^ 2) -\frac {M _ 1M _ 2M _ 3}{R ^ 2} = 0, \\[12pt]
\displaystyle a _ 1M _ 1 + a _ 2M _ 2 + a _ 3M _ 3-D = 0, \\[12pt]
\displaystyle
2 (\alpha _ 1\frac {(M _ 3-M _ 2)} {M _ 1} + \alpha _ 2\frac {(M _ 1-M _ 3)} {M
_ 2} + \alpha _ 3\frac {(M _
2-M _ 1)}{M _ 3} + \\[12pt]
\displaystyle (\beta _ 1M _ 1 + \beta _ 2M _ 2 + \beta _ 3M _ 3) \frac {1} {R ^
2} = 0, \\[12pt]
\end{array}
\end{equation}
where $D$ is the value of integral of the moment. The given system can be solved
numerically. Results of numerical research of bifurcational diagrams
for cases of various values of intensities are presented on~Fig.~6.

{\bf Remark 6.} {\it As well as in case of a plane we are interesting
only in positive roots of the equations~(\ref{Vr14}).
Besides we should notice, that range of values of functions, determined by
integrals of energy and moment are limited by values
$ D _{m} $ and $ E _ {m} $}
$$ D _{m} = 4R ^ 2\frac {(a _ 1a _ 2 + a _ 2a _ 3 + a _ 1a _ 3) ^ 2} {a _ 1a
_ 2a _ 3}, $$
$$
E _{m} = (4R ^ 2)^{\sum a_k} \prod_k \left ({\frac
{ a _ 1 + a _ 2 + a _ 3}{a _ 1 + a _ 2 + a _ 3 + a _ 1a _ 2a _ 3/a _ k ^ 2}}
\right)^{a _ k},
$$
{\it In the latter case appropriate value of the complete moment is calculated
by the formula}
$$
d_E=(4R^2)(a_1+a_2+a_3)\sum_k {\frac {a _ k}
{(a_1+a_2+a_3+a_1a_2a_3/a_k^2)}}.
$$

As initial approach for the solutions of the equations (\ref{Vr14})
with small values of the complete moment we chose the roots of the
equations (\ref{Tp15}). Continuation of bifurcational curves with increase
of the moment (mutual distances) was made by predictor-corrector  method
with use of the Newton iterative procedure.

For each set of intensities we give the bifurcational diagrams and
plots of relative and absolute motions, on which we show the value of angular
speed, corner of an inclination of a plane of vortices under the relation
to an axis of rotation, square of eigenvalue, that is determining the
stability of linearized system.

Interesting effect of a spherical motion of vor\-te\-ces, absent in flat case, is
the birth of new (and in a considered case steady) collinear con\-fi\-gu\-ra\-tions
from a problem of two vortices by values $d_i=4R^2(a_j+a_k)$. There is a dis\-in\-te\-gra\-tion of one vortex of
total intensity $\Gamma_i+\Gamma_j$ to two vortices with intensities
$\Gamma_i$ and $ \Gamma_j$ with increase of $D.$ As shown in Fig.~6, these
configurations aspire to merge with collinear configurations, turning out
with con\-ti\-nu\-a\-tion on parameter
from a flat problem (with small $ D/R ^ 2 $), and then cease to exist, with
further increase of $ D.$

Let's consider bifurcational diagram for a case $a _ 1\ne a _ 2\ne a _ 3\ne
a_1$ (see Fig.~6a). The Thomson's configuration (top curve) exist
only by energy $ E\le E_{T}.$ With $ D = d _ T $, by reaching maximum
possible value $ E_{T}<E_{m}$, Tomson's configuration merges
(point A) with collinear configuration, determined by the most bottom
branch (appropriate to bottom branch on Fig.~2 bifurcational
diagram for plane). Thus all three vortices lay in an equatorial plane, but distance between
them are not equal each other. Passing through a maximum of energy $E_{m}$
by $D=d_E$, this configuration evolve futher with increases of $D$ to a
problem of two vortices (with $D=d_3$). Two other
collinear configurations appropriate to collinear curves on Fig.~2 for
the plane, merge with collinear solutions, arising from
problems of two vortices with values of the moment $d_1, d_2$
(disintegration of one vortex).
Such arising configurations are absent in case of a plane.
The occurrence could be understand from Fig.~8, that shown take off
physical area from borders, on that the moment achieved value $M_k=0.$

In a case, if the ratio $a_1=a_2\ne a_3$ is true, there is a merge not only
Thomson's and collinear branch in a point $A$, but also collinear
configurations, arising by $D=d_1=d_2$ (see. Fig.~6b).

At last for three equal intensities all possible
configurations merge together in a point $A$ by the maximal values
energy $E=E_m=E_T$ and moment $D=d_T.$

In an absolute motion with increase of the mo\-ment (mutual distances)
angular speed of rotation of Thomson's configurations monotonously
decreases (see~Fig.~10), and corner of an inclination normal of a triangle to an
axis of rotation grows monotonously also from zero (as for a plane) up to value
$\pi/2$ by the moment of merge of Thomson's and collinear configurations
(Fig.~9), by exception of a case equal intensities, when an inclination of an
axis change not.

{\bf Remark 7.} {\it The only possible motions of stationary
configurations (i.e. configurations, with distance between vortices do not
varies) are the rotations around of some motionless axis.
It is a consequence of representability of the equations of vortical dynamics
as the  first order equations relatively to dynamic variables.}

In contrast of Thomson's configuration, axis of rotation of collinear
configurations always lays in a plane of vortices and does not vary by
change values of the moment.

With increasing of the moment the diagrams of angular speed
are slumped monotonously and disappearance with merging.
It is necessary to note large size of angular speeds of vortices, arising at the
moment of birth of a new vortex from a problem two  vortices. Within the
framework of the accepted model this increase of angular speed concerns
to different trajectories, but if is present "weak" dissipation, that it's
possible observe slow evolution both values of energy and moment. Certainly,
there are the additional condition existing of such effects in a physical
situation more important is the stability of appropriate stationary
motions.

Part of collinear configurations, occurring from similar configurations on a plane,
 are unstable already in linear approach, as it
is visible from Fig.~11, because they being analogues configurations on a plane.
However collinear configurations, appearing from a problem two vortices,
are steady. The nature of this stability is well visible from geometrical
interpretation presented on Fig.~8. It is possible, that the phenomena of such
grades occurring in an atmosphere of the Earth (that have obviously
dissipation). That is crucial for occurrence of various catastrophic processes
(type hurricanes), accompanying sharp reorganization of dynamics of vortical
formations.  Inverse process of collaps (merge) of two vortices,
that is impossible for model of an ideal liquid, in case of small dissipation
and reduction $ D $ can result in formation of atmospheric vortices with
large angular speed of rotation.

The Thomson's solutions are steady up to the moment of passage through
static configuration. The analogue of the formula (\ref{Tpf}) for Thomson's
configurations on sphere is the expression
\begin{equation}
\label{Tpg}
\lambda ^ 2 = \frac {D-3R^2(a _ 1 + a _ 2 + a _ 3)}{9D^2} a_1a_2a_3
(a _ 1 + a _ 2 + a _ 3).
\end{equation}
That shows, that Tomson's configuration is unstable with value the moment
$$ D> 3R ^ 2 (a _ 1 + a _ 2 + a _ 3),$$
appropriate to the maximal value, for such configurations.

For a case of one negative intensity
$ (\Gamma _ 1 < 0, \quad -\Gamma _ 1> \Gamma _ 2 + \Gamma _ 3) $ bifurcational
diagram is given on~Fig.~6d. In this case behaviour
bifurcational curves by increase $ D $ is similar to behaviour by already
considered situations. Existing in case of a plane Thomson's and
collinear configurations merge in a point $A$ (see Fig.~6d), and after this
disappear by merging from one of collinear branches, that birth from a problem t
of two vortices. Difference from a case only positive by increase $D$
intensities is displayed in existence of collinear solutions, not
limited on energy from above. These solutions occur also
due to stationary configurations of a problem of two vortices.
The change of parameters of an absolute motion of vortices is qualitative by
nothing differs from a case positive intensity except for that, that all of
collinear configuration are steady, in that time as Thomson's --- is unstable
(Fig.~11).

In summary we shall explicitly allocate the basic differences of a spherical
case from flat, arising with increase of the complete moment $ D: $
\begin{enumerate}
\item arising of  Thomson's and collinear configurations on sphere;
\vspace{-3mm}
\item arising of  steady collinear configurations from a problem of two
 vortices rotating with large (infinite) angular speed in the moment of
 occurrence;
\vspace{-3mm}
\item an inclination and evolution of a plane Thomson's configurations;
\vspace{-3mm}
\item existence of static configurations on sphere.
\end{enumerate}
\vspace{-2mm}
Researches of other opportunities of a motion of three vortices on a plane and
to sphere resulting to collaps and which is running up to trajectories will be
are given in the following part of work.

The authors thank I.S.Mamaev and N.N.Simakov for useful discussions and help in work.
The work is carried out under the support of Russian Fond of Fundamental Research
(96--01--00747) and Federal programm "States Support of Integration High Education
and Fundamental Science" (pro\-ject~No.~294).

\begin{figure}[ht!]
$$
\includegraphics{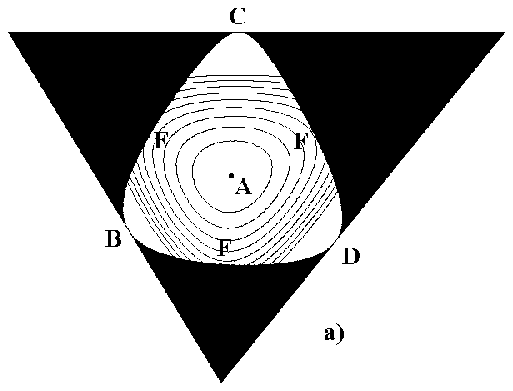}
\includegraphics{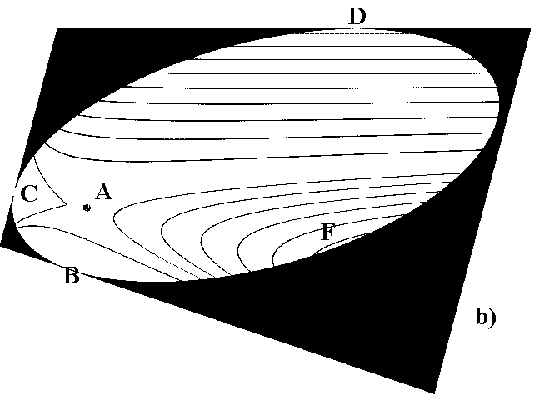}
$$
\caption{Dark part corresponds to nonphysical area
on a plane, that is set by linear integral $ D = const $ and limited
by conditions $ M _ k> 0, $ for a compact case:
a) All of intensities are positive;
b) One of intensities is negative.
The point $A$ corresponds to the Tomson's solution; $F$ ---
collilear solutions; $B, C, D$ --- points, in with which two vortices are
merged in one.}
\end{figure}

\begin{figure}[ht!]
$$
\includegraphics{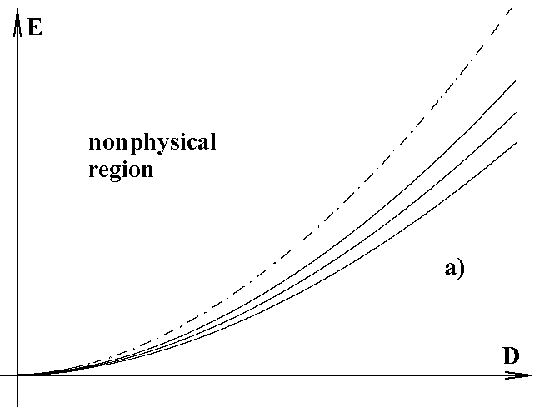}
\includegraphics{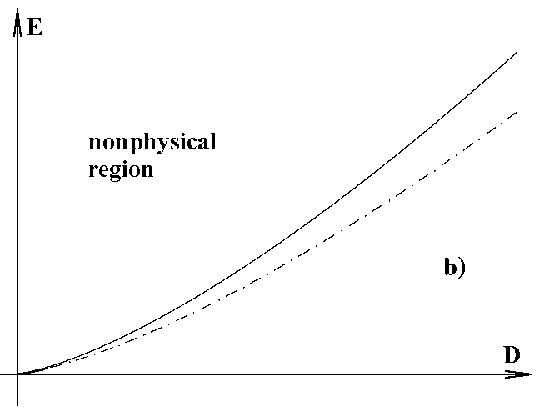}
$$
\caption{The view of bifurcation diagram for three vortices on a plane
in compact case:
a) With positive intensities;
b) One negative intensity. The dot-dashed line corresponds to the Thomson{'}s
solution, continuous lines --- to collinear configurations.}
\end{figure}

\begin{figure}[ht!]
$$
\includegraphics{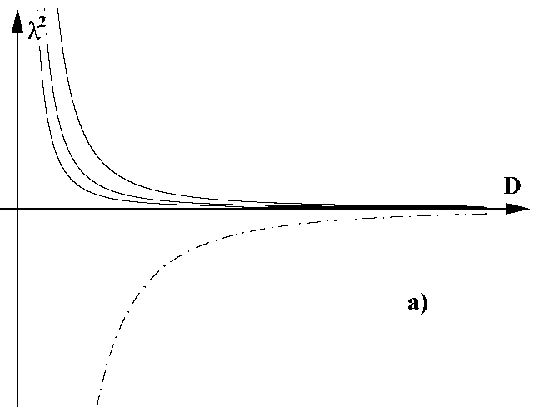}
\includegraphics{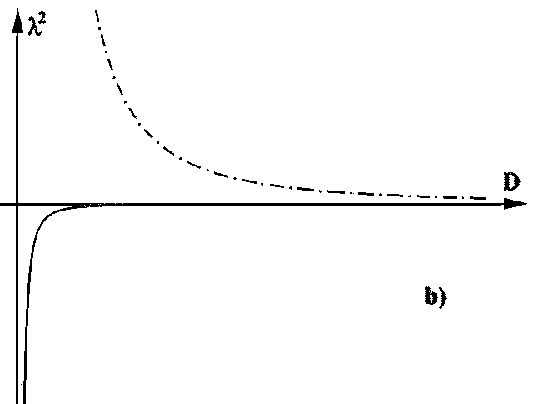}
$$
\caption{Dependence of a squares of  eigenvalues,
determining stability in linear approach from integral of complete
moment for stationary configurations in a compact case:
a) With positive intensities;
b) One negative intensity.
The dot-dashed line corresponds to the Thomson's solutions,
continuous lines --- to collinear configurations.}
\end{figure}

\begin{figure}[ht!]
\begin{center}
\begin{tabular}{cc}
\includegraphics{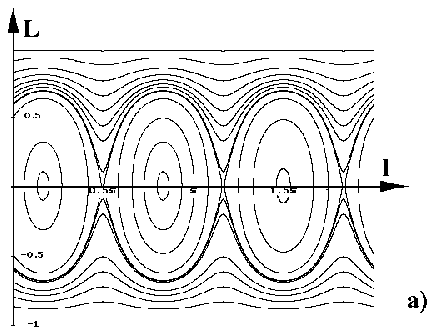} & \includegraphics{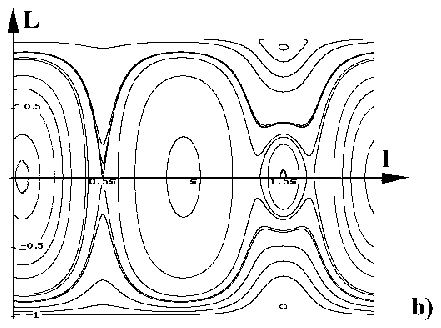} \medskip\\
\includegraphics{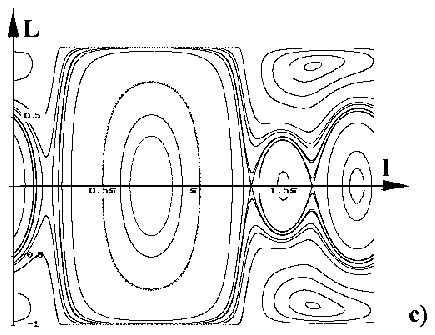} & \includegraphics{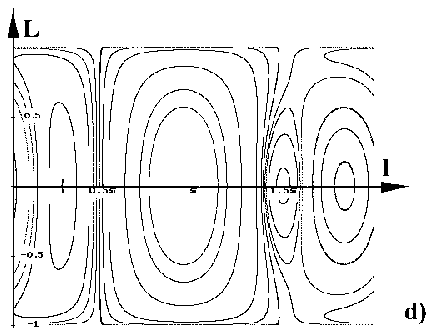}\\
\end{tabular}
\end{center}
\caption{Phase plane for cases:
a) Positive equal intensities;
b) Case $ a _ 1 = a _ 2 \ne a _ 3 $
c) Three positive various intensities;
d) One negative intensity.}
\end{figure}

\begin{figure}[ht!]
$$
\includegraphics{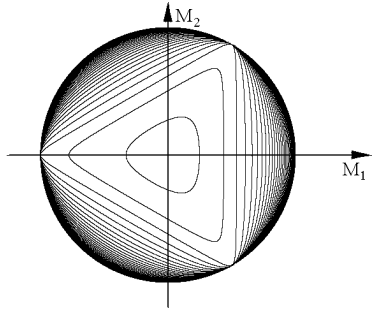}
$$
\caption{Deformation of section of symplectic sheet
for linearized algebra of vortices on sphere depending on change
of sphere's radius of curvature.}
\end{figure}

\begin{figure}[ht!]
\begin{center}
\begin{tabular}{cc}
\includegraphics{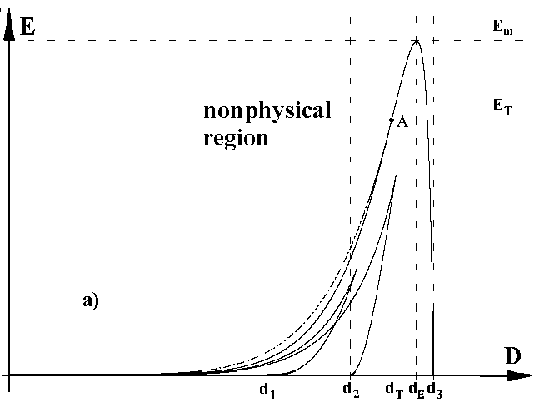} & \includegraphics{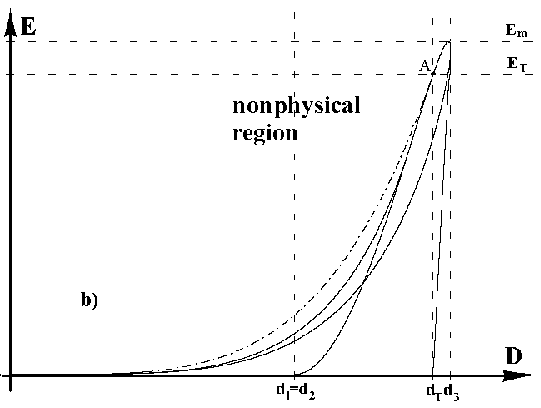} \medskip\\
\includegraphics{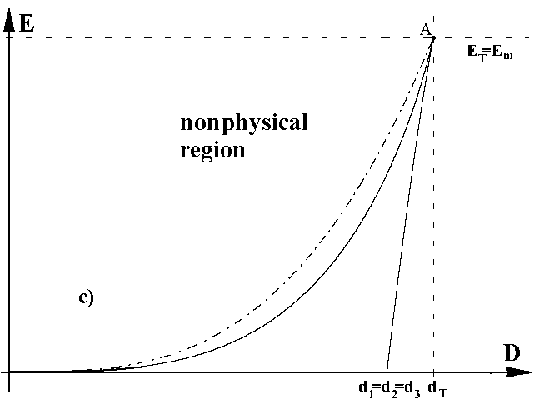} & \includegraphics{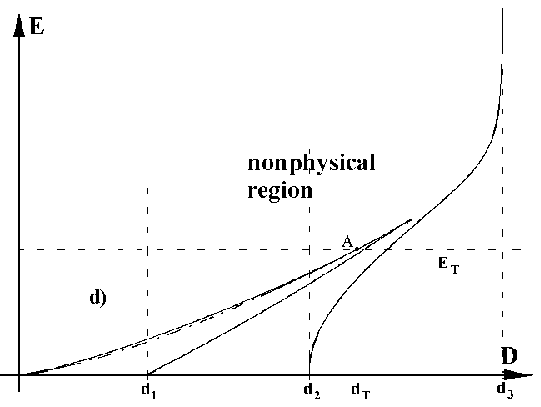}\\
\end{tabular}
\end{center}
\caption{The bifurcational diagram for cases:
a) Various positive intensities;
b) Positive intensities (two coincide);
c) Equal positive intensities;
d) One negative intensity.
The designations of curves coincide with Fig.~2.}
\end{figure}

\begin{figure}[ht!]
\begin{tabular}{c}
\includegraphics{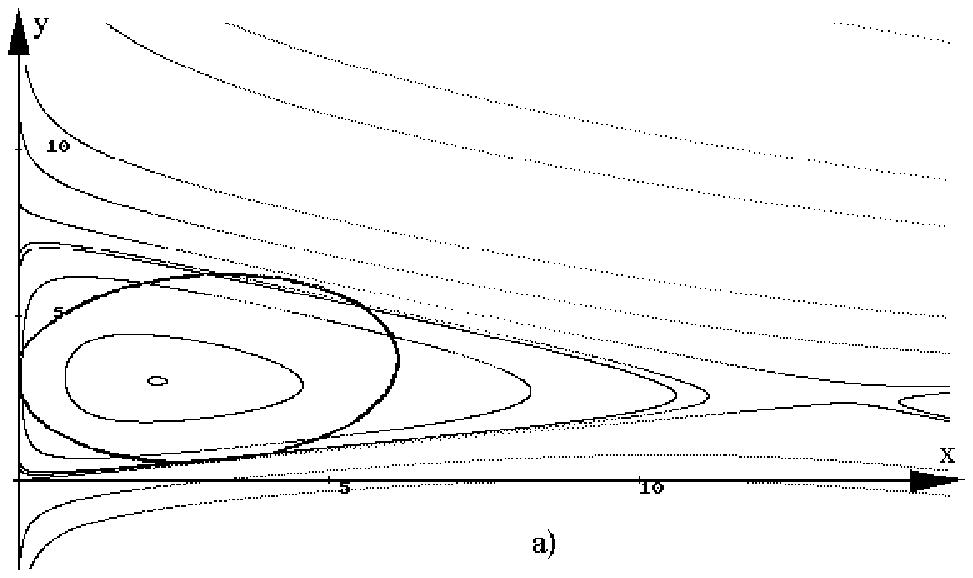}\\
\includegraphics{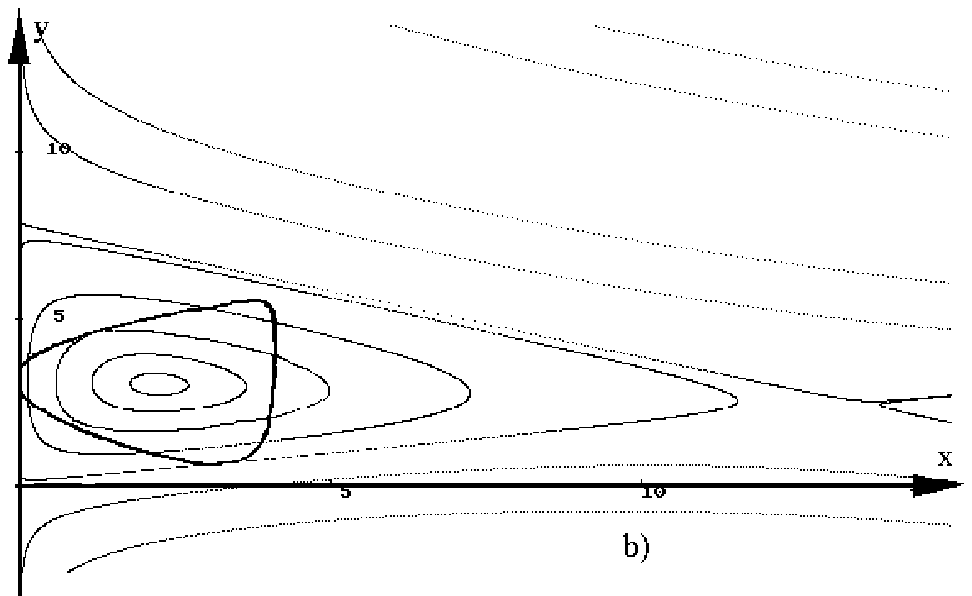}\\
\end{tabular}
\caption{Geometrical picture is isomorfic with a Lotky-Volterra
problem. A part of a phase plane of Lotky-Volterra, that is
limited by a contour, corresponds to physical area of a problem of three
vortices: a) on a plane; b) on sphere with a set of identical intensities.}
\end{figure}

\begin{figure}[ht!]
$$
\includegraphics{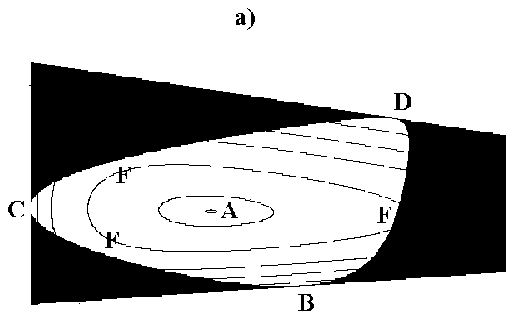}
\includegraphics{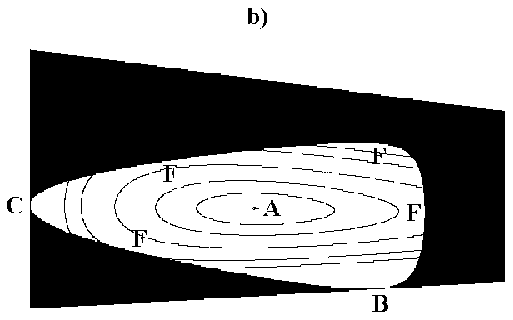}
\includegraphics{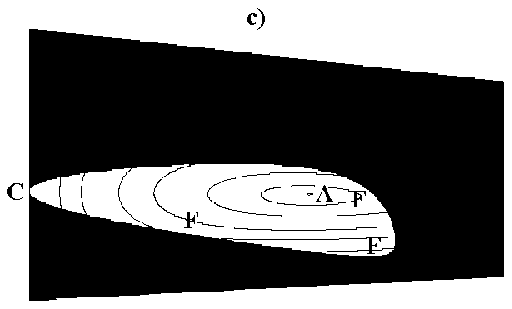}
$$
\caption{Geometrical picture of birth of collinear solutions
from a problem of two vortices. Consecutive figures a), b), c) show change
of points of a contact of borders of physically of allowable area
with increase of the complete moment.
The designations correspond Fig.~1.}
\end{figure}

\begin{figure}[ht!]
\begin{center}
\begin{tabular}{cc}
\includegraphics{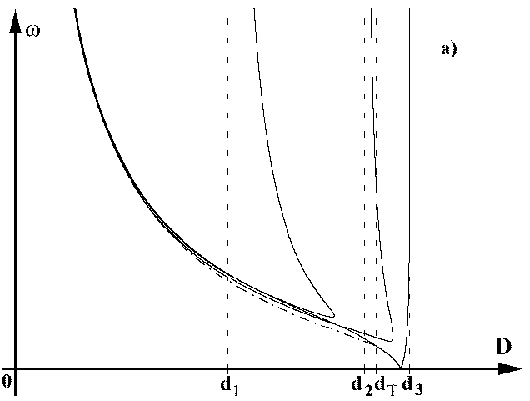} & \includegraphics{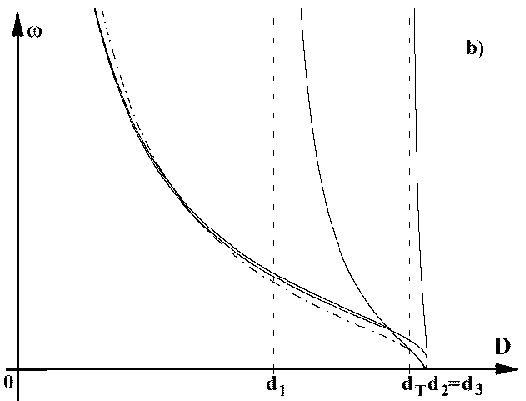} \medskip\\
\includegraphics{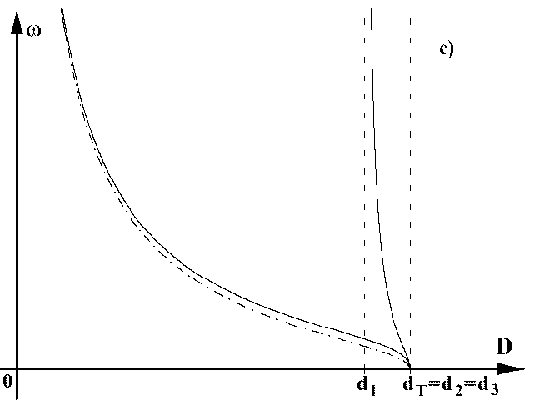} & \includegraphics{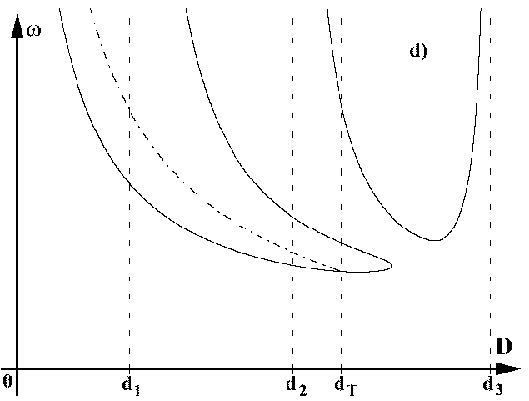}\\
\end{tabular}
\end{center}
\caption{Angular speeds for cases:
a) Various positive intensities;
b) Positive intensities (two coincide);
c) Equal positive intensities;
b) One negative intensity.}
\end{figure}

\begin{figure}[ht!]
$$
\includegraphics{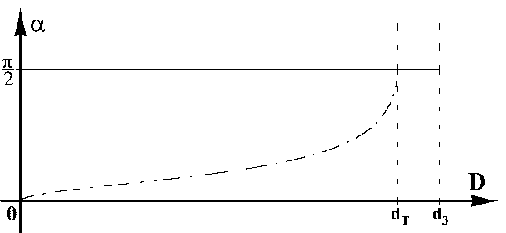}
\includegraphics{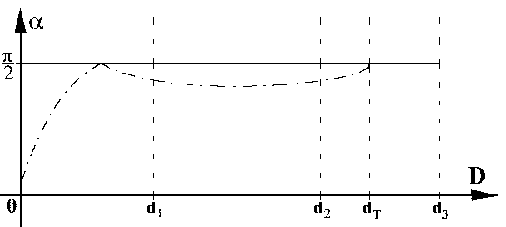}
$$
\caption{Corners of an inclination for cases:
a) Various positive intensities;
b) One negative intensity.}
\end{figure}

\begin{figure}[ht!]
\begin{center}
\begin{tabular}{cc}
\includegraphics{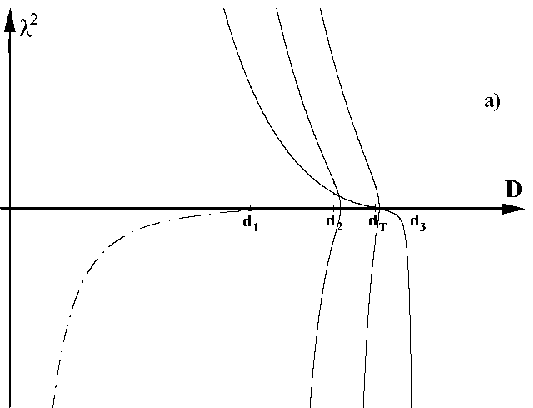} & \includegraphics{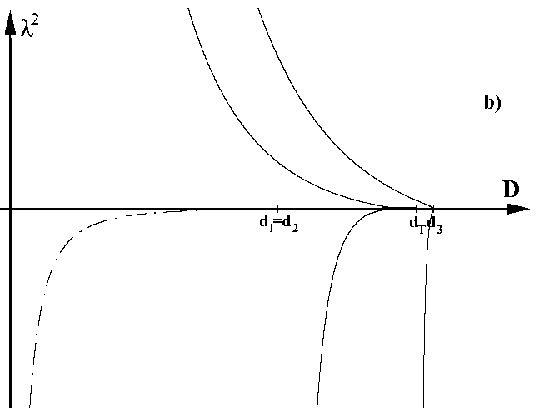} \medskip\\
\includegraphics{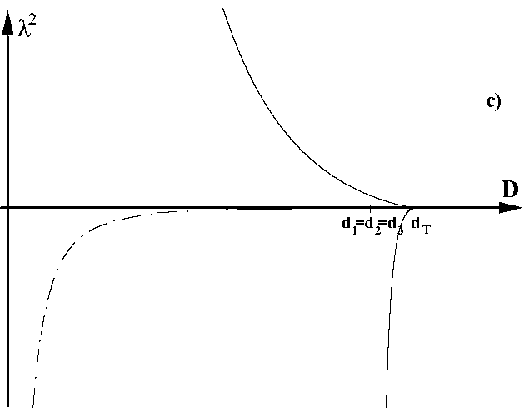} & \includegraphics{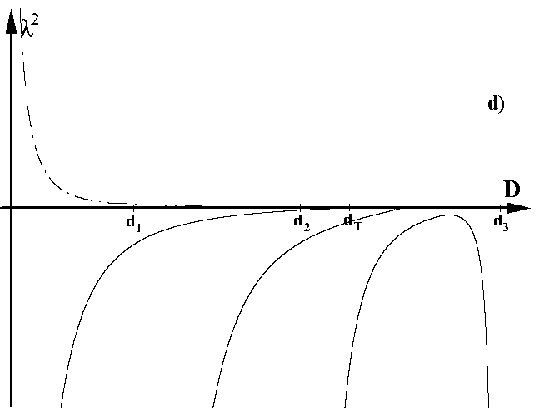}\\
\end{tabular}
\end{center}
\caption{Factor of stability  $\lambda^2$ for cases:
a) Various positive intensities;
b) Positive intensities (two coincide);
c) Equal positive intensities;
d) One negative intensity.}
\end{figure}

\end{document}